\documentclass[12pt]{article}
\usepackage{amssymb}
\usepackage{amsmath}

\usepackage{amstext}
\usepackage{graphicx,epsfig}
\usepackage{epsfig}
\usepackage{verbatim} 
\usepackage{fancyhdr}
\usepackage{fancybox}
\usepackage{color}
\usepackage{ulem,bbold}
\usepackage{enumitem}
\usepackage{subfigure}
\usepackage{bbm}
\usepackage{parskip}
\usepackage{cite}
\usepackage{cite}
\newcommand{\la}{\langle}
\newcommand{\ra}{\rangle}

\newcommand{\Comment}[1]{{}}
\definecolor{MyDarkBlue}{rgb}{0.15,0.15,0.45}
\usepackage[linktocpage=true]{hyperref}
\hypersetup{
colorlinks=true,
citecolor=MyDarkBlue,
linkcolor=MyDarkBlue,
urlcolor=MyDarkBlue,
}

\newcommand{\be}{\begin{equation}}
\newcommand{\ee}{\end{equation}}
\newcommand{\bea}{\begin{eqnarray}}
\newcommand{\eea}{\end{eqnarray}}
\newcommand{\beas}{\begin{eqnarray*}}
\newcommand{\eeas}{\end{eqnarray*}}
\newcommand{\nn}{\nonumber}

\newcommand{\half}{\frac{1}{2}}

\numberwithin{equation}{section}

\begin{document}


\begin{center}
{\Large \bf{Hidden Conformal Symmetry of the\\}}
\vspace{.2cm}
{\Large \bf{ Discrete Series Scalars in dS$_2$ }}
\end{center}

\vspace{1truecm}
\thispagestyle{empty}
\centerline{\Large Kara Farnsworth,${}^{\rm a,}$\footnote{\href{mailto:kmfarnsworth@gmail.com}{\texttt{kmfarnsworth@gmail.com}}} Kurt Hinterbichler,${}^{\rm b,}$\footnote{\href{mailto:kurt.hinterbichler@case.edu} {\texttt{kurt.hinterbichler@case.edu}}} Samanta Saha,${}^{\rm b,}$\footnote{\href{mailto:sxs2638@case.edu} {\texttt{sxs2638@case.edu}}} }

\vspace{.5cm}

\centerline{{\it ${}^{\rm a}$D\'{e}partment de Physique Th\'{e}orique,}}
\centerline{{\it Universit\'{e} de Gen\`{e}ve, CH-1211 Gen\`{e}ve, Switzerland}} 
\vspace{.25cm}

\centerline{{\it ${}^{\rm b}$CERCA, Department of Physics,}}
\centerline{{\it Case Western Reserve University, 10900 Euclid Ave, Cleveland, OH 44106}} 
\vspace{.25cm}

\begin{abstract}

In $D$ dimensional de Sitter space, a scalar field has an infinite tower of special tachyonic mass values at which enhanced shift symmetries appear.  After modding out by these shift symmetries, these fields correspond to the unitary irreducible representations of the de Sitter group known as the discrete series.  
We show that in $D=2$ these theories have global conformal symmetry.  In all but the massless case, these theories have no stress tensor and the conformal symmetry does not act in the usual way on the scalar field.   We find the conformal symmetry by explicitly computing the correlators of the shift invariant local operators and showing that they take conformally invariant forms.  We also demonstrate how these fields are self dual in $D=2$, and dual to the shift invariant massive vector fields, which are therefore also conformally invariant.

\end{abstract}

\newpage

\thispagestyle{empty}
\tableofcontents

\setcounter{page}{1}
\setcounter{footnote}{0}

\parskip=5pt
\normalsize

\section{Introduction\label{introsec}}

On $D$ dimensional de Sitter space (dS$_D$) with radius 
$1/H$, a massive scalar field $\phi$ exhibits shift symmetries for the specific masses 
\bea
  m_k^2= -k(k+D-1)H^2\,,\ \ \ k=0,1,2,\ldots\ \  ,\label{shifmendaee}
\eea
where the integer $k$ is called the level of the shift symmetry \cite{Bonifacio:2018zex}. 
The value $k=0$ corresponds to the massless scalar field, and the higher $k$'s are all tachyonic. The explicit form of the shift transformation for level $k$ is
\bea
\delta \phi = S_{A_1\cdots A_{k}} X^{A_1}\cdots X^{A_{k}}\,. \label{shiftsymmee}
\eea
Here $X^A(x)$ is the standard embedding of the dS$_D$ space into an auxiliary flat Lorentzian ambient space of dimension $D+1$ with metric $\eta_{AB}=\rm{diag}(-1,1,\ldots,1)$, where the dS hyperboloid is the surface $\eta_{AB}X^AX^B =1/H^2$, and $S_{A_1\cdots A_{k}}$ is a constant ambient space tensor which is fully symmetric and fully traceless.

Despite their tachyonic masses, these fields correspond to unitary irreducible representations of the de Sitter group, known as the discrete series scalars (see  \cite{Dobrev:1977qv,Boers:2013pba,Basile:2016aen,Sun:2021thf,Sengor:2022lyv,Sengor:2022kji,Enayati:2022hed,RiosFukelman:2023mgq,Schaub:2024rnl} for reviews of the representation theory of the de Sitter group).
This correspondence occurs once the shift symmetries \eqref{shiftsymmee} are modded out, i.e. gauged.  This gauging removes the would-be unstable modes of the tachyonic mass, leaving a stable unitary theory.  The unitarity can be seen field theoretically by noting that these theories arise as the decoupled longitudinal modes of massive higher spin fields as the higher spin fields approach their maximal depth partially massless \cite{Deser:1983mm,Deser:2001us,Deser:2001pe,Deser:2003gw} values from above \cite{DeRham:2018axr,Bonifacio:2018zex}.  Since the maximal depth partially massless value corresponds to the Higuchi bound \cite{Higuchi:1985ad,Higuchi:1986py}, this limit is unitary.

These shift symmetric scalars make an appearance in various diverse and seemingly unrelated contexts.  The massless case, $k=0$, has been extensively studied due to its potential importance in the cosmology of the early and late universe, see e.g. \cite{Allen:1985ux,Allen:1987tz}.  Other aspects of this case and its quantization are discussed in e.g. \cite{Vilenkin:1982wt,Starobinsky:1982ee,Allen:1987tz,Folacci:1992xc,Joung:2007je,Marolf:2008it,Bros:2010wa,Epstein:2014jaa} (see \cite{Sengor:2023buj} for recent work on the possible cosmological effects of the other $k$ values.)
The $k = 1$ case has emerged in studies related to the conformal factor in quantum gravity \cite{Antoniadis:1991fa,Folacci:1996dv,Faizal:2011iv,Mora:2012zi,Morrison:2013rqa,Miao:2013isa}. 
 In $D=2$, it can be seen explicitly that it arises when gravity is coupled to a conformal matter field with large central charge in dS$_2$ (see footnote 6 of \cite{Anninos:2021ene} and discussions in \cite{Folacci:1996dv,Anninos:2023lin}).
As discussed recently in \cite{Anninos:2023lin}, the discrete series scalars in $D=2$ are realized in a topological $SL(N,\mathbb{R})$ BF theory (at least at the level of the pre-Hilbert space, namely the states that exist before projecting to the gauge invariant states).  The case $N=2$ is related to de Sitter gravity \cite{Isler:1989hq,Chamseddine:1989yz}.
Finally, another completely different context in which these field appear is the following: the fluctuations in the entanglement entropy in a $D$ dimensional conformal field theory (CFT) across a spatial $D-1$ sphere as the state fluctuates away from the vacuum is described by a $k=1$ scalar field propagating on the auxiliary kinematic space of spatial spheres, which is a dS$_D$~\cite{deBoer:2015kda,deBoer:2016pqk} (the $k\geq 2$ cases describe a similar situation involving higher spin conserved currents).

These shift symmetries are seemingly unrelated to the $\frak{so}(2,D)$ global conformal symmetry of the dS$_D$ space itself (as opposed to the $\frak{so}(1,D)$ conformal symmetry of the boundary, which comes from the isometries of dS$_D$).
On dS$_D$, there is a well known conformal mass value where the scalar field acquires the full $\frak{so}(2,D)$ conformal symmetry of dS$_D$,
\be  m_{\rm conformal}^2={D(D-2)\over 4} H^2\,.\label{confomassvaleueee}\ee
At this mass value, conformal symmetry acts in the usual linear, local manner on $\phi$, and there are no other mass values for which this is true (see \cite{Farnsworth:2024iwc} for a recent exposition in our conventions).  For $D>2$, this conformal mass squared is positive, and so it does not coincide with any of the shift symmetric values in \eqref{shifmendaee}.  For $D=2$ on the other hand, the conformal mass value is zero, which corresponds to the $k=0$ shift symmetric value.

In this paper, we will see that the remaining shift symmetric fields in $D=2$, i.e. those with $k\geq 1$ in \eqref{shifmendaee}, are also conformally invariant, and hence define CFTs living on dS$_2$.  This conformal symmetry is hidden, in the sense that it does not act in the standard local manner on $\phi$, though it is still linearly realized.  This can happen because once the shift symmetries are gauged, $\phi$ is no longer a proper gauge invariant operator in the theory, but is more like a gauge field.   The gauge invariant operators are field strengths made from combinations of derivatives of $\phi$ which are invariant under the shift symmetries \eqref{shiftsymmee}.  We will find the conformal symmetry of these theories by computing the two-point correlators of the appropriate field strengths and seeing that they have the unique structure dictated by conformal symmetry. Thus the conformal symmetry acts in the standard manner on the gauge invariant field strengths, which behave as primaries under conformal transformations, though not on the gauge field $\phi$.  Since these theories are free, the conformal symmetry of the two-point correlators is sufficient to establish conformal symmetry for the whole theory (up to possible anomalies at coincident points).

For $k\geq 1$, the removal of the non-shift invariant operators by gauging the shift symmetry also removes the stress tensor, so these new CFTs do not have a well-defined stress tensor.  They have the structure of generalized free CFTs built on conserved traceless symmetric tensor fields.

This same phenomenon -- conformal symmetry that appears at the level of correlators of gauge invariant operators but not in the usual way in the action -- also occurs for linearized gravity (i.e. the free massless spin-2 field) in $D=4$, on both flat \cite{Farnsworth:2021zgj} and curved \cite{Farnsworth:2024iwc} space: the action is not invariant when the linearized metric is transformed in the usual way as a conformal field, but correlators of the   Weyl tensor (which are invariant under the gauge symmetries -- linearized diffeomorphisms in this case) take the proper conformal form.  The same happens for all the higher spins $>2$ in $D=4$, whereas the spin-1 case, i.e. Maxwell electromagnetism, has conformal symmetry in $D=4$ in the usual way acting on the action, and therefore also the gauge invariant correlators \cite{El-Showk:2011xbs}.  Like the $D=2$ scalars with $k\geq 1$, the massless higher spin theories for $s\geq 2$ do not have a gauge invariant stress tensor. 
In these senses, the level $k$ shift symmetric scalar in $D=2$ is a toy model for a massless spin $k+1$ field in $D=4$; in particular, the $k=1$ case is a toy model for linearized gravity.

In $D=2$, the shift symmetric scalars have a self duality \cite{Hinterbichler:2024vyv}, analogous to electromagnetic duality in $D=4$ and generalizing the T-duality of the massless scalar in $D=2$.  
This duality stems from the fact that in $D=2$ the scalar discrete series representations each split into two irreducible halves (see \cite{Joung:2007je,Kitaev:2017hnr,Chen:2022wpa} for reviews of the representation theory of the dS group in $D=2$).  In addition, a massive scalar is dual to a massive vector in $D=2$, and the shift symmetric scalars for $k\geq 1$ are dual to the shift symmetric vectors.  
Here we will be able to see all these dualities explicitly at the level of the correlators of gauge invariant operators.  The conformal symmetry of the shift symmetric scalars then implies that all the shift symmetric vectors also have a hidden conformal symmetry, in the same hidden manner as the $k\geq 1$ scalars.

This paper is structured as follows: In Section \ref{reviewdssec}, we review conformal symmetry for scalar fields in de Sitter space and the shift symmetric fields.  In Section \ref{confd2sec} we explicitly demonstrate that the correlators of the shift invariant field strengths have the conformal structure.   
In Section \ref{dualitiessec} we demonstrate the self duality of the shift symmetric scalar fields in $D=2$ and their duality with the shift symmetric vector fields, showing that the shift symmetric vector fields are also conformal in dS$_2$.  We conclude in Section \ref{conclusionssec} and discuss some open problems.

\textbf{Conventions:}  We work on de Sitter space with radius $1/H$, so that the Riemann tensor is given by $R_{\mu\nu\rho\sigma}=H^2\left(g_{\mu\rho}g_{\nu\sigma}-g_{\mu\sigma}g_{\nu\rho}\right)$ (the curvature conventions are those of \cite{Carroll:2004st}).  The spacetime dimension is $D$ and the metric signature is mostly plus.  The notation $(\cdots)_T$ means to take the fully symmetric and fully traceless projection of the enclosed indices, e.g. $t_{(\mu\nu)_T}=\half \left(t_{\mu\nu}+t_{\nu\mu}\right)-{1\over D}g_{\mu\nu}t^\rho_{\ \rho}$.

When working with correlators we use Euclidean de Sitter, which is the sphere of radius $1/H$.  We make use of stereographic projective coordinates $x^\mu$ in which the metric becomes 
\be g_{\mu\nu}=\Omega^2(x) \eta_{\mu\nu}\,,\ \ \ \Omega(x) \equiv {1\over 1+{H^2\over 4} x^2}\,,   \label{dsmetrice}\ee
where $x^2\equiv \eta_{\mu\nu}x^\mu x^\nu$ and $\eta_{\mu\nu}={\rm diag}(1,\ldots,1)$ is the flat Euclidean metric.  In these coordinates, conformal flatness is manifest, as is the flat space limit $H\rightarrow 0$. 

We often use the variable $Z= \cos (H\mu)\in[-1,1]$ with $\mu$ the geodesic distance between $x$ and $x'$, so that $Z = 1$ ($\mu = 0 $) is the coincident point on the sphere and $Z = -1$ ($\mu = \pi/H$) is the antipodal point.  In the coordinates \eqref{dsmetrice}, $Z$ takes the form
\be Z=1-{H^2\over 2} \Omega(x)\Omega(x') (x-x')^2\,.\label{conformalepforZe}\ee

In a CFT on dS$_D$, the form of the two point functions of conformal primaries at separated points are completely fixed: for a spin $s$ primary ${\cal O}_{\mu_1\cdots\mu_s}$ of scaling dimension $\Delta$, in the coordinates \eqref{dsmetrice} the two point function becomes the following up to an overall constant (see \cite{Farnsworth:2024iwc} for details):
\be \langle {\cal O}_{\mu_1\cdots\mu_s}(x){\cal O}_{\mu'_1\cdots\mu'_s}(x')  \rangle_{\rm dS}=\Omega(x)^{-\Delta+s}\Omega(x')^{-\Delta+s}\langle {\cal O}_{\mu_1\cdots\mu_s}(x){\cal O}_{\mu'_1\cdots\mu'_s}(x')  \rangle_{\rm flat}   \, ,\label{weylrelatedcorree}\ee
where
\be \langle {\cal O}_{\mu_1\cdots\mu_s}(x){\cal O}^{\mu'_1\cdots\mu'_s}(x')  \rangle_{\rm flat} = {1\over |x-x'|^{2\Delta}}I_{(\mu_1}^{(\mu'_1}\cdots I_{\mu_s)_T}^{\mu'_s)_T},\ \ \ I_{\mu\mu'}\equiv \eta_{\mu\mu'}-2{(x-x')_{\mu} (x-x')_{\mu'}\over (x-x')^2}\,, \label{Idefiniteee}\ee
is the unique conformal structure for two point functions on flat space \cite{Osborn:1993cr}.
\section{Scalar fields on dS$_2$ and conformal symmetry\label{reviewdssec}}

Consider a canonical free scalar $\phi$ of mass $m$ in dS$_D$.  The Lagrangian is
\bea
\mathcal{L} = \sqrt{-g} \left(- \frac{1}{2}\nabla_\mu \phi \nabla^\mu \phi - \frac{1}{2} m^2 \phi^2 \right)\,.
\label{actions}
\eea
The standard form of a conformal transformation on $\phi$ is
\bea
\delta \phi = -\left[\xi^\mu \nabla_\mu +\frac{\Delta}{D} \nabla \cdot \xi  \right]\phi\,,\label{confosymms}
\eea
where $\Delta$ is a real number giving the scaling dimension of $\phi$ and $\xi^\mu$ is a conformal Killing vector of dS$_D$, i.e. a vector field satisfying the conformal Killing equation $\nabla_{(\mu}\xi_{\nu)_T}=0$ (see \cite{Farnsworth:2024iwc} for details).

The Lagrangian \eqref{actions} has \eqref{confosymms} as a symmetry for only one value of $m^2$ and $\Delta$, given by
\bea 
m^2_{\rm conformal}= \frac{D(D-2)}{4} H^2 , \:\:\:\: \Delta= \frac{D}{2}-1\,.
\label{massterm}
\eea 
In particular, in $D=2$ the theory is conformally invariant under \eqref{confosymms} only for the massless case $m^2=0$, with scaling dimension $\Delta = 0$.

\subsection{Correlator}

Consider the scalar correlator $\langle  \phi(x)\phi(x')\rangle$.  For simplicity we work on Euclidean dS$_D$, which is just the sphere of radius $1/H$.  By dS symmetry, the correlator depends only on the geodesic distance between $x$ and $x'$.   Away from the coincident point $x=x'$, it must be regular and must be annihilated by the Klein-Gordon operator $\nabla^2 -m^2$.  At the coincident point, it must have the same singularity as the flat space correlator.  These requirements fix the correlator to be (see \cite{Allen:1985wd, Farnsworth:2024iwc} for details)
\bea
\langle \phi(x)\phi(x')\rangle =  \frac{H^{D-2}\Gamma(\delta_{+})\Gamma(\delta_{-})}{2^D \pi^{D/2}\Gamma(D/2)} \, _2F_1\left(\delta_{-},\delta_{+};\frac{D}{2};\frac{Z+1}{2}\right),
\label{twopointcorrelator}
\eea
where
\bea
\delta_{\pm} \equiv \frac{D-1}{2}\pm \sqrt{\frac{(D-1)^2}{4}-\frac{m^2}{H^2}}\,.
\label{eq:deltascalar}
\eea
Here $Z$ is the variable defined in the conventions at the end of Section \ref{introsec}.

For the conformal mass value \eqref{massterm} when $D>2$, the correlator reduces to a simple form which upon using the coordinates \eqref{dsmetrice} can be seen to be the form \eqref{weylrelatedcorree} compatible with conformal symmetry \cite{Farnsworth:2024iwc}.

\subsection{Discrete series values}

The correlator \eqref{twopointcorrelator} is well defined as a function of $m^2$ except when $m^2$ takes the values
\be
 m^2_k = -k(k+D-1)H^2,\:\:\:\:\:\:\: k = 0,1,2,3,\ldots, 
 \label{polesvaee}
\ee
where we have $\delta_+=d+k$, $\delta_-=-k$.
At these mass values  there is a simple pole (coming from $\Gamma(\delta_{-})$).  These are precisely the mass values corresponding to the scalar discrete series representations for dS$_D$, and it is at these values that the theory \eqref{actions} acquires the Galileon-like extended shift symmetry \eqref{shiftsymmee}.

Despite the fact that the discrete series mass values are tachyonic ($m_k^2<0$) for $k\geq 1$, these representations are all unitary.  Realizing these unitary representations requires gauging the shift symmetry (as discussed in e.g. \cite{Folacci:1992xc}).  At the level of local operators, this means that we consider only local operators that are invariant under the shift symmetries \eqref{shiftsymmee}.  The basic such operator is the symmetric and traceless ``field strength'' tensor $F_{\mu_1\cdots\mu_{k+1}}$, which is constructed from the $(k+1)$-th symmetrized traceless derivative \cite{Bonifacio:2018zex},
\bea
F_{\mu_1\cdots \mu_{k+1}} \equiv \nabla_{(\mu_1} \cdots \nabla_{\mu_{k+1})_T}\phi\,. \label{fieldstrengtss}
\eea
Although the correlator \eqref{twopointcorrelator} is divergent, if we regularize the divergence, the correlators of $F_{\mu_1\cdots \mu_{k+1}}$ will be finite as the regulator is removed.  Since all other invariant operators are constructed from derivatives and powers of this operator modulo the equations of motion \cite{Hinterbichler:2024vyv}, it is these finite correlators which define the discrete series theories.  We will see explicitly how this works for $D=2$ in what follows.

\section{Conformal symmetry of the discrete series in $D=2$\label{confd2sec}}

In this section we demonstrate that the discrete series scalar theories, as defined above, have global conformal symmetry in $D=2$.

In $D = 2$ the two point correlation function (\ref{twopointcorrelator}) for the massive scalar becomes,
\bea
\langle \phi(x) \phi(x')\rangle  =  \frac{1}{4 \pi} \Gamma(\delta_{+})\Gamma(\delta_{-}) \, _2F_1\left(\delta_{-},\delta_{+};1;\frac{Z+1}{2}\right),
\label{d=2twopoint}
\eea
with, 
\be
\delta_{\pm} \equiv \frac{1}{2}\pm \sqrt{\frac{1}{4}-\frac{m^2}{H^2}} \,.
\ee
This is well defined except for the shift symmetric mass values \eqref{polesvaee}
\be
 m_k^2 = -k(k+1)H^2,\:\:\:\:\:\:\: k = 0,1,2,3,\ldots, 
 \label{poles}
\ee
where we have $\delta_+=k+1$, $\delta_-=-k$.  In these cases there are simple poles coming from the $\Gamma(\delta_-)$.

\subsection{$k=0$}

We start with the case $k=0$, the massless scalar, which is already well known to be conformal in $D=2$: the is the one case in which the conformal mass value \eqref{massterm} overlaps with one of the shift symmetric values \eqref{polesvaee}.

In this case the field is massless, $m^2=0$, and has the simple shift invariance under the transformation  
\be
\phi(x)\rightarrow \phi(x)+ S\,, 
\label{shifttransformation}
\ee 
where $S$ is a constant.  Gauging this shift symmetry, the basic field strength operator \eqref{fieldstrengtss} is a vector operator, given by taking a single derivative of $\phi$,
\be
F_\mu =\nabla_\mu \phi\,.  
\ee

Let us see that the correlator of this shift invariant operator is finite. 
Starting with the correlator \eqref{d=2twopoint}, we define the Green's function $G_0$ by regularizing and isolating the divergence by setting $\delta_-\rightarrow \epsilon$, $\delta_+\rightarrow 1-\epsilon$   
and then expanding for small $\epsilon$. We get\footnote{We find the package HypExp \cite{Huber:2005yg} useful for Taylor expanding the hypergeometric functions.}
\bea
G_0(Z) \equiv \frac{1}{4\pi\epsilon}-{1\over 4\pi}\log(1-Z)+\textnormal{const.}+{\cal O}(\epsilon)\,.\label{2ptde2e}
\eea
The divergence is now isolated as a $1/\epsilon$ pole.  Different ways of regularizing may change the coefficient of this pole or the value of the additive constant finite part, but the coefficient of the logarithm is regulator independent.  Due to this logarithmic dependence, the expression \eqref{2ptde2e} is not conformally invariant (despite the fact that in this case the action is conformally invariant under the standard conformal transformation \eqref{confosymms}).  
To form the correlator of $F_\mu$, we take a derivative of \eqref{2ptde2e} at each of the points $x$ and $x'$,
\bea
\langle F_\mu(x)F_{\mu'}(x')\rangle = \nabla_\mu \nabla_{\mu'}G_0(Z),
\label{correlatorF_mu}
\eea
where $\nabla_{\mu'}$ indicates the covariant derivative with respect to $x'$. 
Plugging in \eqref{2ptde2e}, the derivatives can be evaluated using the rules in \cite{Allen:1985wd}, and the result is
\be
\langle F_\mu(x)F_{\mu'}(x')\rangle = \frac{H^2}{4\pi (1-Z) } \left(g_{\mu\mu'} + 2 n_\mu n_{\mu'} \right).
\label{correlatorsk=0}
\ee
Here $g_{\mu\mu'}(x ,x')$ is the parallel propagator along the geodesic between $x$ and $x'$, $n_\mu(x , x') $ is the unit normalized tangent vector to this geodesic at $x$, pointing away from $x'$, and  $n_{\mu'}(x, x') $ the analogous normal at $x'$ pointing away from $x$. (See Appendix A of \cite{Farnsworth:2024iwc} for more details on this setup and a summary of the rules for taking derivatives of these quantities, in our conventions).

The result \eqref{correlatorsk=0} is finite as expected (the $1/\epsilon$ pole is a constant and is killed by the derivative), and is regulator independent (the regulator dependent additive constant in the finite part is killed by the derivative).  The interesting fact for us is that this correlator for $F_\mu$ is now conformally invariant: it has the correct structure for a two-point function of a spin-1 primary of scaling dimension $\Delta=1$.  To see this, we rewrite it in terms of the conformally flat coordinate system \eqref{dsmetrice}.  Using \eqref{conformalepforZe} and similar rules for the other quantities (see Appendix A of \cite{Farnsworth:2024iwc} for a summary) the correlator \eqref{correlatorsk=0} becomes
\bea
\langle F_\mu(x)F_{\mu'}(x')\rangle =\frac{1}{2\pi |x-x'|^2} \left(\eta_{\mu\mu'} -2\frac{(x-x')_\mu (x-x')_{\mu'}}{(x-x')^2}\right).
\label{2pointk=0case}
\eea
This has the  conformally invariant structure \eqref{weylrelatedcorree} for $s=1$ and $\Delta=1$ (so that the factors of $\Omega$ in \eqref{weylrelatedcorree} vanish).   

For $D>2$, the correlator of $F_\mu$ will still be finite, but it will not have the conformal structure.

Gauging the shift symmetry thus gives a CFT where the primary operator is a vector with weight 1.  This is well known, and is the same story as that of the massless scalar in $D=2$ flat space (as presented in e.g. chapter 2 of \cite{Polchinski:1998rq}), where the scalar correlator has logarithms that break conformal symmetry but the correlators of the derivative of the scalar are conformal.  Below we will see that a similar story occurs for the other discrete series scalars, though with the crucial difference that the conformal symmetry is not seen in the usual way in the action.

\subsection{$k=1$\label{k1sec}}

According to (\ref{poles}), for $k = 1$ in $D=2$ the scalar field has a mass $m^2 =  - 2 H^2$. This theory is invariant under the shift transformation of the scalar field realized by the ambient space Galileon-like transformation
\be
\delta \phi = S_A X^A\,.
\ee
The invariant field strength \eqref{fieldstrengtss} is then a symmetric traceless rank-2 tensor, made from two derivatives of the field,
 \be
 F_{\mu\nu} = \nabla_{(\mu} \nabla_{\nu)_T}\phi \,.\label{k1fieldsstrenghte}
 \ee

We define the Green's function $G_1$ by regularizing and isolating the divergence in the scalar two point function \eqref{d=2twopoint} via $\delta_-\rightarrow -1+\epsilon$, $\delta_+\rightarrow 2-\epsilon$ and expanding for small $\epsilon$, 
{\bea
G_1(Z) \equiv \frac{Z}{4\pi\epsilon}- \frac{ Z}{4\pi} \log(1-Z)+\left({\rm degree} \ 1 {\rm \ polynomial \ in\ } Z\right)+{\cal O}(\epsilon)\,.\label{k1loggree}
\eea}
Different choices of regularization may change the coefficient of the $1/\epsilon$ pole, or the coefficients of the polynomial in the finite term, but the coefficient of the log will be regulator independent.

The correlator of $F_{\mu\nu}$ is then
\be
\langle  F_{\mu\nu}(x) F_{\mu' \nu'}(x')\rangle= \nabla_{(\mu} \nabla_{\nu)_T} \nabla_{(\mu'} \nabla_{\nu')_T}G_1(Z)\,,
\ee
which can be computed using the rules described in Appendix A of \cite{Farnsworth:2024iwc}.  The result can be written as
\begin{align}
\langle  F_{\mu\nu}(x) F^{\mu' \nu'}(x')\rangle = &\ \frac{3H^4}{2 \pi  (1-Z)^2} \left[\tilde I_{(\mu}^{(\mu'} \tilde I_{\nu)_T}^{\nu')_T} - (1+Z) \left( g_{[\mu}^{\ \ \mu^\prime}g_{\rho}^{\ \ \nu^\prime}n_{\nu]} n^\rho+ 2 g^{\mu^\prime [\rho^\prime}g_\nu^{\ \ \alpha^\prime}  n^{\nu^\prime]} n_{\rho^\prime} g_{\mu \alpha^\prime} \right)\right] ,  \nn\\ \label{f2ptk1exeee}  \\
& \tilde I_{\mu\mu'}\equiv  g_{\mu \mu'}+2n_\mu n_{\mu'}\,.
\end{align}
The $1/\epsilon$ pole has cancelled and the result is finite and regularization independent, depending only on the structure of the log term in \eqref{k1loggree} and its coefficient.

The part of \eqref{f2ptk1exeee} proportional to $\tilde I_{(\mu}^{(\mu'} \tilde I_{\nu)_T}^{\nu')_T}$ gives precisely the conformally invariant form \eqref{weylrelatedcorree} for a spin $s=2$ conformal primary of weight $\Delta=2$.  The rest of \eqref{f2ptk1exeee} does not have this conformally invariant form, however it vanishes identically in $D=2$ because it involves antisymmetrizing over more than two indices.  In the conformally flat coordinates \eqref{dsmetrice}, and using \eqref{conformalepforZe} and the similar rules for the other quantities in Appendix A of \cite{Farnsworth:2024iwc} (which gives $\tilde I_{\mu\mu'}=\Omega(x)\Omega(x') I_{\mu\mu'}$ with $ I_{\mu\mu'}$ as defined in \eqref{Idefiniteee}),
the correlator \eqref{f2ptk1exeee} becomes
\bea
\langle F_{\mu\nu}(x) F^{\mu' \nu'}(x')\rangle = \frac{6}{\pi}
\frac{1}{|x-x'|^4} I_{(\mu}^{(\mu'}  I_{\nu)_T}^{\nu')_T}\,  . \label{FFcorrelatfineee}
\eea

We thus see that, in $D=2$, the $k=1$ discrete series scalar defines a CFT on dS$_2$.  It is a generalized free CFT built on the symmetric traceless primary operator $F_{\mu\nu}$.  Note that $\Delta=2$ saturates the unitarity bound for a spin-2 operator in $D=2$, so $F_{\mu\nu}$ is a conserved operator, $\nabla^\nu F_{\mu\nu}=0$, as can be seen upon plugging in \eqref{k1fieldsstrenghte} and using the Klein-Gordon equation for $\phi$, or directly by taking the divergence of the correlator \eqref{f2ptk1exeee}.

Note that, unlike the $k=0$ case, the $k=1$ theory does not have a well-defined stress tensor.  For $k=0$, the stress tensor involves only derivatives of the field, so it is shift invariant, can be written in terms of the field strength $F_\mu=\nabla_\mu\phi$, and is included as an operator in the theory.  For $k=1$, the canonical stress tensor is no longer invariant under the extended $k=1$ shifts (and cannot be made so by improvements), so it cannot be expressed in terms of $F_{\mu\nu}$ and is not an operator in the theory.  Note that $F_{\mu\nu}$ itself is a symmetric traceless conserved operator with $\Delta=2$, just like a stress tensor; but it is not a stress tensor because it does not generate the conformal transformations: to generate the conformal transformations it would have to have a non-vanishing three point function, but it is linear in $\phi$ and so its three point function is zero.

There is an analogy between what we have seen for the $k=0$ and $k=1$ scalars in $D=2$ on the one hand, and what occurs for electromagnetism and linearized gravity in $D=4$ on the other hand.  Electromagnetism is conformally invariant only in $D=4$: the action is conformally invariant in the standard way and the correlators of the gauge invariant Maxwell field strength tensor have the conformally invariant structure \cite{El-Showk:2011xbs}.   
Also, the stress tensor can be made gauge invariant by improvement and is expressible in terms of the field strength, and thus it exists as an operator in the theory.   All of these statements have analogs in the $k=0$ scalar theory in $D=2$.  Linearized gravity is also conformally invariant only in $D=4$: the correlators of the linearized Weyl tensor (the basic gauge invariant field strength, analogous to the Maxwell field strength of electromagnetism) have the conformally invariant structure \cite{Farnsworth:2021zgj}.  However, unlike electromagnetism, the action is not conformally invariant in the usual way. 
There is no stress tensor, because the canonical stress tensor cannot be made gauge invariant through improvements.  This situation is analogous to the $k=1$ scalar theory in $D=2$.

\subsection{General $k$}

We now show that the discrete series scalars for all $k$ are conformal in $D=2$.  The calculation below is a Euclidean version of a similar calculation which can be found in Appendix A.3 of \cite{Loparco:2023rug}.  We define the Green's functions $G_k$ by regularizing and isolating the divergence in the scalar two point function \eqref{d=2twopoint} via $\delta_-\rightarrow -k+\epsilon$, $\delta_+\rightarrow 1+k-\epsilon$ and expanding for small $\epsilon$,
\bea
G_k(Z)   \equiv \frac{1}{4\pi \epsilon } {P}_k (Z)-{1\over 4\pi} {P}_k (Z) \log\left (1-Z\right)+\left({\rm degree} \ k {\rm \ polynomial \ in\ } Z\right)+{\cal O}(\epsilon). \label{gdnkzexpe}
\eea
Here, ${P}_k(Z)$ is the $k$-th Legendre polynomial of $Z$, which is a degree $k$ polynomial in $Z$.  Different choices of regularization may change the coefficient of the $1/\epsilon$ pole, or the coefficients in the polynomial in the finite term, but the coefficient of the log will be regulator independent. From this, we want to compute the correlator of the field strength \eqref{fieldstrengtss},
\be \langle F_{\mu_1\cdots \mu_{k+1}} F_{\mu'_1\cdots \mu'_{k+1}} \rangle= \nabla_{(\mu_1} \cdots \nabla_{\mu_{k+1})_T}  \nabla_{(\mu'_1} \cdots \nabla_{\mu'_{k+1})_T}  G_k(Z) .\ee

At this point, we can greatly simplify the computation by working directly in the conformally flat coordinates \eqref{dsmetrice}, and in addition use complex coordinates on the Euclidean plane.  Taking the Cartesian coordinates on the plane to be $x^\mu=(x,y)$, we define the usual complex coordinates,
\be
z= x+iy, \:\:\: \:\: \Bar{z}=x-iy \, . \ \ \ 
\ee
The ordinary derivates are then
\be
\partial\equiv \partial_z = \frac{1}{2}(\partial_x- i\partial_y),\:\:\:\: \Bar{\partial}\equiv \partial_{\Bar{z}} = \frac{1}{2}(\partial_x+ i\partial_y)\,.
\ee
The metric on Euclidean dS$_2$ now reads,
\be
g_{\mu\nu}= \Omega^2(z,\bar{z})\left(\begin{array}{cc}
 0 & \frac{1}{2}  \\
 \frac{1}{2} & 0 \\
\end{array}\right),\:\:\: g^{\mu\nu}= \Omega^{-2}(z,\bar{z})\left(\begin{array}{cc}
 0 & 2  \\
 2 & 0 \\\end{array}\right)\, ,\ \ \  \Omega(z,\bar{z})= \frac{1}{1+\frac{H^2}{4}z\Bar{z}}\,. \label{complexmetrice}
\ee

The field strength \eqref{fieldstrengtss} is a symmetric traceless tensor.  As such, it only has two independent components in $D=2$.
Those two components are the one with all $z$ indices and the one with all $\Bar{z}$ indices (all the mixed components contribute only to the traces, as can
be seen from the fact that the metric \eqref{complexmetrice} is off-diagonal).  In terms of the $(x,y)$ coordinates these are
\bea
    &&F \equiv F_{z\cdots z}= \frac{1}{2}(F_{x\cdots x}-iF_{x\cdots xy})\, , \ \ \  \Bar{F} \equiv F_{\Bar{z}\cdots\Bar{z}} = \frac{1}{2}(F_{x\cdots x}+iF_{x\cdots xy})\,. \label{FFbardefee}
\eea
Since the field strength has only two components, there are only three correlator components we need to compute,
\bea
&&\langle F(z,\Bar{z}) F(z',\Bar{z}')\rangle = \nabla^{k+1}\nabla^{\prime^{k+1} } G_k(Z)\,,\nn\\
 &&   \langle \Bar{F}(z,\Bar{z}) \Bar{F}(z',\Bar{z}')\rangle = \Bar{\nabla}^{k+1}\Bar{\nabla}'^{k+1}G_k(Z)\,,\nn\\
&&   \langle F(z,\Bar{z}) \Bar{F}(z',\Bar{z}')\rangle= \nabla^{k+1}\Bar{\nabla}'^{k+1} G_k(Z)\,. \label{ddFFexpredxe}
\eea
Here, we use the covariant derivatives in complex coordinates, $\nabla \equiv \nabla_z,\ \nabla' \equiv \nabla_{z'},\ \Bar{\nabla} \equiv \nabla_{\Bar{z}},\ \Bar{\nabla}' \equiv \nabla_{\Bar{z}'}$.  Another nice thing about the conformal complex coordinates is that the only nonzero Christoffel symbols are those will all $z$'s or all $\bar z$'s,
\be \Gamma^{z}_{zz}=-{H^2\over 2}\Omega\,\bar z\, ,\ \ \ \Gamma^{\bar z}_{\bar z\bar z}=-{H^2\over 2}\Omega \, z\, .\ee
When acting on a tensor with $s_z$ of the $z$ indices and $s_{\bar z}$ of the $\bar z$ indices, the covariant derivatives thus take the form
\bea
    \nabla= \partial - s_z \frac{H^2}{2}\Omega \Bar{z} ,\:\:\:\:\:\: \Bar{\nabla}= \Bar{\partial} - s_{\bar z} \frac{H^2}{2}\Omega z\,. \label{covdiparcommplee}
\eea

In terms of $z$, $\Bar{z}$, the variable $Z$ that the correlators depend on can be written as
\bea
    Z= 1- \frac{H^2}{2} \frac{1}{(1+\frac{H^2}{4}z\Bar{z})} \frac{1}{(1+\frac{H^2}{4}z'\Bar{z}')} (z-z') (\Bar{z}- \Bar{z}')\,. \label{Zinterdnze}
\eea
Now, from direct computation using \eqref{covdiparcommplee}, we can see that all the repeated complex derivatives of $Z$ vanish,
\bea
    \nabla^2 Z = \Bar{\nabla}^2 Z = \nabla'^2 Z= \Bar{\nabla}'^2 Z =0 \,.
\eea
Since we can never have more than one covariant derivative of one type hitting $Z$, we have, for any function $f(Z)$,
\bea
    \nabla^{k} f(Z)= f^{(k)}(Z) (\partial Z)^{k}\,,\ \ \  \bar\nabla^{k} f(Z)= f^{(k)}(Z) (\bar\partial Z)^{k}\,, \label{multzderivGee}
\eea
where $f^{(k)}(Z)$ is the $k$-th derivative of $f(Z)$ with respect to $Z$. Now the field strength correlators now can be calculated in terms of ordinary derivatives of the Green's function \eqref{gdnkzexpe} for the level $k$ scalar,
\begin{align}
  \langle F(z,\Bar{z}) F(z',\Bar{z}')\rangle &= \nabla^{k+1}\nabla'^{k+1} G_{k}(Z) \, \nn \\
 &=G_{k}^{(2k+2)}(Z) (\partial Z)^{k+1} (\partial' Z)^{k+1} +\cdots + (k+1)!G_{k}^{(k+1)}(Z)(\partial \partial' Z)^{k+1} \,.
\label{DOFofgeneralsoltion}
\end{align}
The expression for $\langle \bar F\bar F\rangle$ is obtained by taking $\partial\rightarrow \bar\partial$, $\partial'\rightarrow \bar\partial'$ and the expression for $\langle F\bar F\rangle$ is obtained by taking $\partial'\rightarrow \bar\partial'$.

From \eqref{DOFofgeneralsoltion}, we can immediately see that the result after plugging in \eqref{gdnkzexpe} for $G_k(Z)$ will be finite and regularization independent, depending only on the structure of the log and its coefficient: the $1/\epsilon$ pole and the non-log finite part are degree $k$ polynomials in $Z$ and there are always more than $k$ derivatives acting on $G_k(Z)$.  
 
 The result for the correlators is (see Appendix \ref{proofappendix} for the proof)
\begin{align} 
\begin{split}
\langle F(z,\Bar{z}) F(z',\Bar{z}')\rangle &= \frac{(-1)^{k+1} (2k+1)!}{4\pi} \frac{1}{(z-z')^{2(k+1)}}\,,\\
\langle\Bar{F}(z,\Bar{z}) \Bar{F}(z',\Bar{z}')\rangle &= \frac{(-1)^{k+1} (2k+1)!}{4\pi} \frac{1}{(\Bar{z}-\Bar{z}')^{2(k+1)}}\, ,\\
  \langle F(z,\Bar{z}) \Bar{F}(z',\Bar{z}')\rangle &= 0\,. 
  \end{split}
   \label{ansddrkzcore}
\end{align}
This is the correct conformal form for the two point function of a traceless symmetric spin $k+1$ primary in $D=2$, with $\Delta=k+1$.   In the $z,\bar z$ coordinates, the general conformal correlator \eqref{Idefiniteee}, with the complex components ${\cal O}(z,\bar z)$, $\bar {\cal O}(z,\bar z)$ of ${\cal O}_{\mu_1\cdots\mu_s}(x)$ defined analogously to \eqref{FFbardefee}, becomes 
\be \la {\cal O}(z,\bar z){\cal O}(0)\ra= {(-1)^s\over 2^s} {1\over (z\bar z)^{\Delta-s}z^{2s}}\, ,\ \  \la \bar{\cal O}(z,\bar z)\bar{\cal O}(0)\ra= {(-1)^s\over 2^s} {1\over (z\bar z)^{\Delta-s}\bar z^{2s}}\, ,\ \  \la {\cal O}(z,\bar z)\bar{\cal O}(0)\ra=0\,, \label{correfifxee}\ee 
and we see that this structure matches \eqref{ansddrkzcore} for $\Delta=s=k+1$.
In the $x^\mu$ coordinates, we thus have 
\be \langle F_{\mu_1\cdots\mu_{k+1}}(x){ F}^{\mu'_1\cdots\mu'_{k+1}}(x')  \rangle = {2^{k-1} (2k+1)! \over \pi}{1\over |x-x'|^{2(k+1)}}I_{(\mu_1}^{(\mu'_1}\cdots I_{\mu_{k+1})_T}^{\mu'_{k+1})_T}\, ,\ \ \ \label{Idefiniteeefee}\ee
with $I_{\mu\mu'}$ as in \eqref{Idefiniteee}.   Note that the complex coordinates automatically take care of all the $D=2$ dimension dependent identities that would be involved in a direct comparison like we did in section \ref{k1sec}.

For $D>2$, the correlators of the field strengths will still be finite, but they will not have the conformal structure.  Thus the scalar shift symmetric fields in higher dimensions are ordinary QFTs rather than CFTs.

$\Delta=s$ saturates the CFT unitarity bound for conformal primaries and is the value for a conserved current.  The field strength tensor is indeed conserved, $\nabla^\mu F_{\mu\mu_2\cdots\mu_{k+1}}=0$, which can be checked directly using the equations of motion or the correlator \eqref{correfifxee}.
In complex coordinates the conservation condition becomes 
\be \bar\partial F=\partial\bar F=0\ .\ee
(These are ordinary derivatives, no connection terms are needed for dS$_2$ in these coordinates.)  This means that $F$ is holomorphic and $\bar F$ is anti-holomorphic, as can be seen from \eqref{ansddrkzcore}.

We see that the level $k$ discrete series scalar with its shift symmetry gauged defines a CFT.  It is a generalized free CFT built on a conserved spin $k+1$, $\Delta=k+1$ primary.
Note that the 2 point function \eqref{Idefiniteeefee} is positive for all $k$, so these discrete series theories are all unitary.

\section{Dualities\label{dualitiessec}}

In $D=2$, the discrete series scalars are self dual and they are dual to massive vector fields \cite{Hinterbichler:2024vyv}.  Here we will see how these dualities work at the level of correlators.
 
\subsection{Self duality}

In $D=2$ Euclidean space we can split symmetric traceless tensors into imaginary self dual and anti-self dual parts: for the level $k$ field strength tensor \eqref{fieldstrengtss} this split is as follows,
\begin{equation}
\begin{gathered}
F^\pm_{\mu_1\cdots\mu_{k+1}}={1\over 2}\left( F_{\mu_1\cdots\mu_{k+1}}\mp i\epsilon^\nu_{\ \mu_1} F_{\nu\mu_2\cdots\mu_{k+1}}\right), \\
 \epsilon^\nu_{\ \mu_1} F_{\nu\mu_2\cdots\mu_{k+1}}^\pm=\pm i \,F_{\mu_1\cdots\mu_{k+1}}^\pm\,,\ \  \ F_{\mu_1\cdots\mu_{k+1}}=F_{\mu_1\cdots\mu_{k+1}}^++F_{\mu_1\cdots\mu_{k+1}}^-\,. 
\end{gathered}
\end{equation}
In terms of the $z,\bar z$ coordinates, we have for the complex components as defined in \eqref{FFbardefee} (note that in these complex coordinates the metric determinant is negative so we use $\epsilon_{\mu\nu}=i{\Omega^2\over 4}\tilde\epsilon_{\mu\nu}$, where $\tilde\epsilon_{\mu\nu}$ is the epsilon symbol with $\tilde\epsilon_{z\bar z}=1$),
\be F^+=0,\ \bar F^+=\bar F,\ \  F^-=F,\ \ \ \ \bar F^-=0,\ee
so $\bar F$ and $F$ are nothing but the self dual and anti-self dual parts of the field strength tensor.

From \eqref{ansddrkzcore}, we have $\langle F \Bar{F}\rangle = 0$ and so the self dual and anti-self dual parts decouple from each other and theory splits into two decoupled sectors.  Furthermore, the self dual sector has holomorphic correlators and the anti-self dual parts have the corresponding anti-holomorphic correlators given by complex conjugation.  This generalizes to $k>0$ the well known holomorphic factorization of the massless scalar CFT in $D=2$.  At the level of representation theory, this factorization is reflected in the fact that the discrete series representations of dS$_2$ each split into two irreducible chiral parts.

\subsection{Duality with the massive vector field}

Consider the Lagrangian for a massive vector field $A_\mu$ with mass $m_A$ on dS$_D$,
\begin{align}
\mathcal{L} = \sqrt{-g}\left( - \frac{1}{4}F^2_{\mu\nu}  - \frac{1}{2} m_A^2 A^2\right).
\end{align}
The correlation function for this theory on Euclidean dS$_D$ can be found in \cite{Allen:1985wd,Narain:2014oja,Belokogne:2016dvd} and can be written as
\begin{align}
\langle A_\mu (x) A_{\mu'}(x')\rangle = f_1(Z) g_{\mu\mu'} + f_2(Z) n_\mu n_{\mu'}\, ,
\label{eq:vector2pt}
\end{align}
with
\begin{align}
f_1(Z) &= \frac{(Z^2-1)}{(D-1)} G_A'(Z) + ZG_A(Z)\,,\\
f_2(Z) &= \frac{(Z^2-1)}{(D-1)} G_A'(Z) + (Z-1)G_A(Z)\,,
\end{align}
and
\begin{align}
G_A(Z) = \frac{H^D}{m_A^2} \frac{(1-D)\Gamma(\delta_+ + 1) \Gamma(\delta_- +1)}{2^{D+1} \pi^{D/2} \Gamma(D/2+1)} \, {}_2 F_1\left(\delta_- +1, \delta_++1; \frac{D+2}{2}; \frac{Z+1}{2}\right),
\end{align}
where 
\begin{align}
\delta_{\pm} = \frac{(D-1)}{2} \pm \sqrt{\frac{(D-3)^2}{4} - \frac{m_A^2}{H^2}}\,.
\label{eq:deltavector}
\end{align}

This correlator is singular in both the massless case (due to the $1/m_A^2$ factor), and at the poles of $\Gamma(\delta_-+1)$, corresponding to the following mass values
\begin{align}
m_A^2 = -(k_A+2)(k_A+D-1)H^2,\quad{} k_A = 0,1,2,\dots\, \ .
\end{align}
These are the mass values for the level $k_A$ shift symmetric vector \cite{Bonifacio:2018zex}.  The invariant field strength for these shift symmetric vectors is a rank $k_A+2$ symmetric traceless tensor,
\be F^{(A)}_{\mu_1\cdots \mu_{k_A+2}}=\nabla_{(\mu_1}\cdots \nabla_{\mu_{k_A+1}} A_{\mu_{k_A+2})_T}\,.\label{AFreelee}\ee

In $D = 2$, $\delta_{\pm}$ of the scalar \eqref{eq:deltascalar} and $\delta_\pm$ of the vector \eqref{eq:deltavector} are equal if $m^2 = m_A^2$.  This reflects the fact that in $D = 2$, a generic massive scalar field is dual to a massive vector field with the same mass $m_A=m$.   This duality is realized via the operator relations 
\be \phi = {1\over m} \epsilon^{\mu\nu} \nabla_\mu A_\nu \,,\ \ \ A_\mu =- {1\over m} \epsilon_{\mu\nu}^{} \nabla^\nu \phi\,,\ \label{scalarvectdualrelee} \ee
in Euclidean dS$_2$, which are consistent with each other upon using the equations of motion.

We can see this duality explicitly from the correlators: taking $D = 2$, the functions $f_1(Z)$ and $f_2(Z)$ in \eqref{eq:vector2pt} simplify to
\begin{align}
\begin{split}
f_1(Z) &= \frac{H^2}{m_A^2} \frac{\Gamma(\delta_+ + 1)\Gamma(\delta_- + 1)}{8\pi }\bigg[2 \, {}_2F_1\left( \delta_- , \delta_+ , 1; \frac{Z+1}{2}\right)  +Z\, {}_2F_1\left( \delta_- + 1, \delta_+ + 1, 2; \frac{Z+1}{2}\right)\bigg], \\
f_2(Z)
&=  \frac{H^2}{m_A^2} \frac{\Gamma(\delta_+ + 1)\Gamma(\delta_- + 1)}{8\pi }\bigg[2\, {}_2F_1\left( \delta_- , \delta_+ , 1; \frac{Z+1}{2}\right)  +(Z+1)\, {}_2F_1\left( \delta_- + 1, \delta_+ + 1, 2; \frac{Z+1}{2}\right)\bigg] .
\end{split}
\end{align}
Now compare the vector correlator $\langle A_\mu(x) A_{\mu'} (x') \rangle $ with the scalar correlation function $\langle \nabla_\mu  \phi(x) \nabla_{\mu'}  \phi(x')\rangle$. In $D = 2$ the latter can be written 
\begin{align}
\nabla_\mu \nabla_{\mu'} \langle \phi(x) \phi(x')\rangle =&\ g_1(Z) g_{\mu\mu'} + g_2(Z) n_\mu n_{\mu'}\,,
\end{align}
with 
\begin{align}
\begin{split}
g_1(Z) =&\ \frac{\Gamma(\delta_++1)\Gamma(\delta_-+1)}{8 \pi} H^2\, {}_2 F_1 \left( \delta_-+1, \delta_++1; 2, \frac{Z+1}{2}\right),\\
g_2(Z) =&\ \frac{\Gamma(\delta_++1)\Gamma(\delta_-+1)}{8 \pi} H^2  \bigg[2\, {}_2F_1\left( \delta_- , \delta_+ , 1; \frac{Z+1}{2}\right)+(1+Z)\, {}_2F_1\left( \delta_- + 1, \delta_+ + 1, 2; \frac{Z+1}{2}\right)\bigg].
\end{split}
\end{align}
Comparing this to the vector, we can now see that we have
\begin{align}
\langle A_\mu(x) A_{\mu'} (x') \rangle 
=&\ \frac{1}{m_A^2}\nabla_\mu \nabla_{\mu'} \langle \phi(x) \phi(x')\rangle-\frac{1}{m_A^2}g_{\mu\mu'} g^{\nu\nu'}\nabla_\nu \nabla_{\nu'}  \langle \phi(x) \phi(x')\rangle \nn \\
=&\ \frac{1}{m_A^2}\epsilon_{\mu\nu}^{} \epsilon_{\mu'\nu'}^{} \nabla^{\nu} \nabla^{\nu'} \langle \phi(x) \phi(x')\rangle\,,\label{dualitychefchee}
\end{align}
where in the last equality we have used the relation $\epsilon_{\mu\nu} \epsilon_{\mu'\nu'} = g_{\mu\mu'} g_{\nu\nu'}-g_{\nu\mu'} g_{\mu\nu'}$
that holds between the epsilon symbols at the two points and the parallel propagator, and we have also used that $g_{\mu\mu'}\nabla^{\mu'}=-\nabla_\mu$, $g_{\mu\mu'}\nabla^{\mu}=-\nabla_{\mu'}$ when acting on a function of $Z$.
The relation \eqref{dualitychefchee} is precisely the duality relation \eqref{scalarvectdualrelee} between the generic massive theories.

The duality between shift symmetric theories now follows.
At the values of the scalar and vector masses where the correlation functions are singular due to shift symmetries, this corresponds to equating
\begin{align}
k = k_A + 1\,.
\end{align}
The correlation functions of the shift invariant operators \eqref{AFreelee} formed out of $A$ at level $k_A$ are equivalent to correlation functions of the operators \eqref{fieldstrengtss} formed out of $\phi$ at level $k_A+1$ (the $k=0$ scalar is not included here and is not dual to any vector theory).  The relation is found by taking symmetrized derivatives of the basic duality relation \eqref{scalarvectdualrelee},
\be F^{(A)}_{\mu_1\cdots \mu_{k+1}}=- {1\over m} \epsilon_{\mu_1}^{\ \ \nu} F^{(\phi)}_{\nu\mu_2\cdots \mu_{k+1}}\,.\ee
 This is reflected in the correlation functions, and as shown above, these are manifestly conformally invariant, so the shift-symmetric vector fields are also conformally invariant.  As with the $k\geq 1$ scalars, this conformal symmetry is seen only in correlation functions of the shift invariant operators and is not manifest in the usual way at the level of the action.

The various dualities and self dualities among the shift symmetric fields in $D=2$ can be summarized as follows:
\bea \raisebox{-35pt}{\epsfig{file=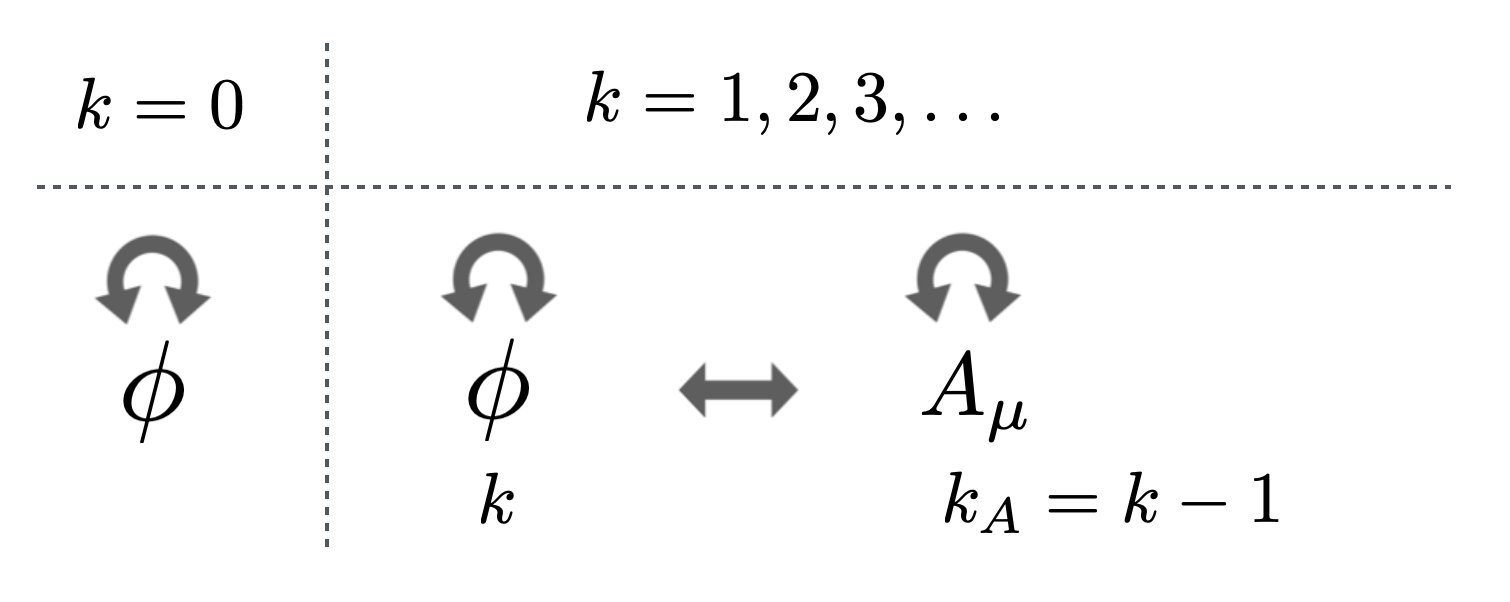,height=1.3in,width=3.2in}}\,\eea
The scalars with $k\geq 1$ are self dual and dual to the vectors with $k_A\geq 0$, which in turn are therefore also self dual.  All of these have the hidden conformal symmetry.  The $k=0$ scalar is dual to itself, and has ordinary conformal symmetry.

\section{Conclusions\label{conclusionssec}}

The shift symmetric scalars are scalar fields on de Sitter space with specific mass values, labelled by a level $k=0,1,2,\ldots$, where extended Galileon-like shift symmetries emerge.  When these shift symmetries are gauged, the fields realize the discrete series representations of the de Sitter group, which are unitary representations despite the fact that the mass values are all tachyonic with $m^2<0$.

Here we have seen that these shift symmetric scalars have global conformal symmetry in dS$_2$.  For $k=0$, this is the standard conformal symmetry of the massless scalar in $D=2$.  For $k\geq 1$, the conformal symmetry is not realized in the standard way at the level of the action, but is realized as conformal symmetry of the correlators of shift invariant local operators.  This is analogous to the conformal symmetry that occurs in linearized gravity.

{The fact that the $D=2$ shift symmetric scalars have conformal symmetry suggests that their representations, the discrete series representations of the dS$_2$ algebra $\frak{so}(1,2)$, should extend to representations of the larger algebra $\frak{so}(2,2)$ of conformal symmetries on dS$_2$.  The branching of representations from $\frak{so}(2,2)\rightarrow \frak{so}(1,2)$ was studied in \cite{Hogervorst:2021uvp,Penedones:2023uqc} and this is indeed the case: see section 5.2 of \cite{Penedones:2023uqc}. }

We also saw explicitly the electromagnetic-like self duality of the $D=2$ shift symmetric scalars, as well as their duality with the shift symmetric vectors.  The self duality is analogous to the self duality of $p$-forms in $2p+2$ dimensions that allows for chiral $p$-forms when $p$ is even.  Lagrangians that propagate chiral $p$ forms are notoriously non-straightforward to construct, and there has been much work on this subject in the past and more recently \cite{Zwanziger:1970hk,Siegel:1983es,Pasti:1995ii,Pasti:1995tn,Pasti:1996vs,Bandos:1997ui,Sen:2015nph,Sen:2019qit,Mkrtchyan:2019opf,Bansal:2021bis,Avetisyan:2022zza,Evnin:2022kqn,Evnin:2023cdf,Hull:2023dgp}.  It should be possible to extend these approaches to construct similar Lagrangians for chiral shift symmetric scalars and vectors in $D=2$.

There is also a connection between the shift symmetric scalars and the higher derivative conformal scalars with a $\Box^n$ kinetic term, which are conformal in the usual way at the level of the action (see Appendix A of \cite{Brust:2016gjy} for a summary of these theories).  As discussed in Appendix D of \cite{Bonifacio:2018zex} (which elaborates on \cite{Baumann:2017jvh}), on de Sitter space these theories can be diagonalized into a sum of ordinary canonical scalars, with alternating signs for their kinetic terms, some of which are shift symmetric:  
For $D\geq 4$ with $D$ even, the $\Box^n$ theory is equivalent to the sum of the first $n+1-D/2$ shift-symmetric scalars, but also contains other non-shift symmetric massive scalars (including the conformal scalar).  For $D$ odd, the $\Box^n$ theory does not contain any of the shift-symmetric scalars.  In $D=2$, however, the $\Box^n$ theory is equivalent to precisely the first $n$ shift symmetric scalars, with no extra massive fields.

As is well known, the massless $k=0$ scalar when $D=2$ has more than just global conformal symmetry, it has a full infinite dimensional Virasoro symmetry generated by the moments of the stress tensor.  We cannot generate a Virasoro symmetry in the same way for the $k\geq 1$ scalars, because they do not have stress tensors.  They could conceivably still have some larger symmetry group that is not generated by local currents, and this would be an interesting question to look into.

Another well known fact about the massless $k=0$ scalar when $D=2$ is that the topology of the field space is important.  Whether the field is ${\mathbb R}$ valued or $U(1)$ valued, and whether the shift symmetry is continuous or discrete, matters for the spectrum of the theory.  This is because there can be other local operators beyond those constructed from polynomials in the field and its derivatives, i.e. the vertex operators and winding operators, whose spectrum depends on these topological considerations.  What the analogous story is for the $k\geq 1$ fields presents many open questions, such as what compactness of the gauge group looks like when the shift transformations have spacetime dependence, what the resulting vertex operators look like, etc.  Some work in this direction in a different context can be found in \cite{Fliss:2021ekk}.

Although we worked in de Sitter space, the same shift symmetries also exist in anti-de Sitter space, with non-tachyonic mass values, and the conclusions about conformal symmetry in $D=2$ will also hold
(see \cite{Blauvelt:2022wwa} for more on the effects of the AdS shift symmetries).

Finally, it would be interesting to understand what happens to the conformal symmetry of the $D=2$ scalars as interactions are included.  Interactions of the shift symmetric theories were studied in \cite{Bonifacio:2018zex,Bonifacio:2021mrf}, and in particular, the $k=1$ scalar has natural self-interactions that deform the symmetry algebra, given by the DBI theory of a 1-brane with dS$_2$ background geometry probing a fixed bulk (A)dS$_3$ geometry \cite{Goon:2011qf,Goon:2011uw}.  This is another point in favor of the interpretation of the level $k$ scalar in $D=2$ as a toy model for massless spin $k+1$ in $D=4$: the $k=0$ theory has natural multi-field non-linear interactions (the non-linear sigma model) analogous to the natural multi-field interactions of the massless spin-1 field (Yang-Mills theory), and the $k=1$ theory has natural non-linear self interactions (the DBI theory) analogous to the natural non-linear self interactions of the massless spin-2 field (general relativity).

\vspace{-5pt}
\paragraph{Acknowledgments:}

We would like to thank Karapet Mkrtchyan, Andy Stergiou and Zimo Sun for discussions and comments.  The work of KF is supported by the Swiss National
Science Foundation under grant no. 200021-205016.  KH acknowledges support from DOE grant DE-SC0009946.

\appendix

\section{Computation of the correlators\label{proofappendix}}

Here we prove that the correlation functions of the field strength components for general $k$ are given by the expressions \eqref{ansddrkzcore}.  The starting point is \eqref{ddFFexpredxe}, with $G_k(Z)$ as given in \eqref{gdnkzexpe} and with $Z$ as given in \eqref{Zinterdnze}.

The $1/\epsilon$ pole and the finite polynomial terms in \eqref{gdnkzexpe} do not contribute because these are at most degree $k$ polynomials and there are always more than $k$ derivatives acting on $G_k(Z)$, so we can take 
{\begin{align}
G_k(Z)  \rightarrow -\frac{1}{4\pi} P_k(Z) \ln(1-Z)\,. \label{Gknopolyee}
\end{align} }
By directly taking covariant derivatives of $Z$ in \eqref{Zinterdnze} with respect to the different complex coordinates, we find the relations 
\begin{align}
\nabla  Z\, {\nabla'} Z &= (1+Z) \nabla  {\nabla'} Z\, ,\label{eq:zz'simp}\\
\nabla  Z\, {\bar\nabla'} Z &= -(1-Z) \nabla  {\bar\nabla'} Z\,, \label{eq:zzbar'simp} \\
\frac{(\nabla  {\nabla'} Z)}{(1-Z)^2}  & = -\frac{1}{2}\frac{1}{(z - z')^2}\, .
\label{eq:Zfunc}
\end{align}

We will now prove by induction that 
\begin{align}
\nabla ^{k+1}{\nabla'}^{k+1} G_k(Z) &= \frac{2^{k+1} (2k+1)!}{4\pi}  \frac{(\nabla {\nabla'} Z)^{(k+1)}}{(1-Z)^{2(k+1)}}\, ,\label{eq:ansatz}
\end{align}
which, upon using \eqref{eq:Zfunc}, immediately gives the first of \eqref{ansddrkzcore}.

We start by establishing the base case $k = 0$.  Using $G_0(Z) \rightarrow -\frac{1}{4\pi}  \ln(1-Z)$,
\begin{align}
\nabla  {\nabla'} G_0(Z) &= \frac{1}{4\pi} \frac{1}{(1-Z)^2}{\nabla'}Z \nabla Z+\frac{1}{4\pi} \frac{1}{(1-Z)}{\nabla'} \nabla Z= \frac{2}{4\pi} \frac{\nabla {\nabla'} Z}{(1-Z)^2}\,,
\end{align}
where in the last equality we used \eqref{eq:zz'simp}. This gives \eqref{eq:ansatz} for $k= 0$.

Now for the induction step we look at $\nabla ^{k+2}{\nabla'}^{k+2} G_{k+1}(Z)$.  Using \eqref{Gknopolyee} and applying Bonnet’s recursion formula for the Legendre polynomials, 
\be (k+1)P_{k+1}(Z)=(2k+1)Z P_{k}(Z)  - k P_{k-1}\, ,\ee
this can be written as
\begin{align}
\nabla ^{k+2}{\nabla'}^{k+2} G_{k+1}(Z) &= \frac{1}{(k+1)} \nabla ^{k+2}{\nabla'}^{k+2} \left[(2k+1)Z G_{k}(Z)  - k \, G_{k-1}(Z)\right] \,,
\end{align}
Because $\nabla ^2 Z = {\nabla'}^2 Z = 0$, this can be expanded as
\begin{align}
\begin{split}
\nabla ^{k+2}{\nabla'}^{k+2} G_{k+1}(Z) =&\  \frac{(2k+1)}{(k+1)}(k+2)^2{\nabla'} \nabla  Z\left({\nabla'}^{k+1} \nabla^{k+1} G_{k}(Z)\right)\\
&+ \frac{(2k+1)}{(k+1)}(k+2) \nabla  Z{\nabla'}\left({\nabla'}^{k+1} \nabla^{k+1} G_{k}(Z)\right)\\
&+ \frac{(2k+1)}{(k+1)}(k+2) {\nabla'} Z\nabla \left({\nabla'}^{k+1} \nabla^{k+1} G_{k}(Z)\right)\\
&+\frac{(2k+1)}{(k+1)} Z{\nabla'} \nabla  \left({\nabla'}^{k+1} \nabla^{k+1} G_{k}(Z)\right)\\
&- \frac{k}{(k+1)} {\nabla'}^2 \nabla ^2 \left({\nabla'}^{k} \nabla^{k} G_{k-1}(Z)\right).
\end{split}
\end{align}

Now plugging in our ansatz \eqref{eq:ansatz} for $\nabla ^{k+1}{\nabla'}^{k+1} G_k(Z) $, this becomes
\begin{align}
\nabla ^{k+2}{\nabla'}^{k+2} G_{k+1}(Z) =&\  \frac{(2k+1)}{4\pi (k+1)} \frac{({\nabla'} \nabla  Z)^{(k+2)}}{(1-Z)^{2(k+2)}}\\
&\ \ \times \bigg[kZ\left( 2^{k+1}(2k+1)!  - 4k(2k+1)2^{k} (2k-1)! \right)\left[kZ+2(k+2)\right]\nonumber\\
&\ \ \ \ \ +\left((k+2)(5k+6) 2^{k+1}  (2k+1)!- 4k^2(2k^2+9k+8)2^{k} (2k-1)! \right)\bigg].\nonumber
\end{align}
The terms in the first line cancel, and this simplifies to 
\begin{align}
\nabla ^{k+2}{\nabla'}^{k+2} G_{k+1}(Z)
=&\  \frac{2^{k+2} (2k+3)! }{4\pi} \frac{({\nabla'} \nabla  Z)^{(k+2)}}{(1-Z)^{2(k+2)}}.
\end{align}
With both a base case and an induction step, this establishes \eqref{eq:ansatz}  and thus the first of \eqref{ansddrkzcore}.

The second of \eqref{ansddrkzcore} follows immediately by taking $z \rightarrow \bar{z}$ in the above.

Finally, to show the third of \eqref{ansddrkzcore}, we note that, using \eqref{eq:zzbar'simp}, 
\begin{align}
\nabla  \bar\nabla' \ln(1-Z) &= - \frac{1}{(1-Z)^2}\nabla Z \bar\nabla' Z- \frac{1}{(1-Z)} \nabla  \bar\nabla' Z= 0.
\end{align}
This means that when taking the derivatives to find our correlation function,
\begin{align}
\nabla^{k+1} \bar\nabla'^{k+1} G_k(Z) &\rightarrow \nabla^{k+1} \bar\nabla'^{k+1} \left[ -\frac{1}{4\pi} P_k(Z) \ln(1-Z) \right],
\end{align}
either all $k+1$ of the $z$ derivatives,  or all $k+1$ of the ${\bar z}'$ derivatives, must hit the Legendre polynomial $P_k(Z)$, but in either case this vanishes because after using \eqref{multzderivGee} we will have $(k+1)$-th derivatives of an order $k$ polynomial.

\renewcommand{\em}{}
\bibliographystyle{utphys}
\addcontentsline{toc}{section}{References}
\bibliography{ds2arxiv}

\providecommand{\href}[2]{#2}\begingroup\raggedright\begin{thebibliography}{10}

\bibitem{Bonifacio:2018zex}
J.~Bonifacio, K.~Hinterbichler, A.~Joyce, and R.~A. Rosen, ``{Shift Symmetries
  in (Anti) de Sitter Space},''
  \href{http://dx.doi.org/10.1007/JHEP02(2019)178}{{\em JHEP} {\bfseries 02}
  (2019) 178}, \href{http://arxiv.org/abs/1812.08167}{{\ttfamily
  arXiv:1812.08167 [hep-th]}}.

\bibitem{Dobrev:1977qv}
V.~K. Dobrev, G.~Mack, V.~B. Petkova, S.~G. Petrova, and I.~T. Todorov,
  \href{http://dx.doi.org/10.1007/BFb0009678}{{\em {Harmonic Analysis on the
  n-Dimensional Lorentz Group and Its Application to Conformal Quantum Field
  Theory}}}, vol.~63.
\newblock 1977.

\bibitem{Boers:2013pba}
M.~Boers, ``{Group theory and de Sitter QFT}: {The concept of mass},'' Master's
  thesis, Groningen U., 2013.

\bibitem{Basile:2016aen}
T.~Basile, X.~Bekaert, and N.~Boulanger, ``{Mixed-symmetry fields in de Sitter
  space: a group theoretical glance},''
  \href{http://dx.doi.org/10.1007/JHEP05(2017)081}{{\em JHEP} {\bfseries 05}
  (2017) 081}, \href{http://arxiv.org/abs/1612.08166}{{\ttfamily
  arXiv:1612.08166 [hep-th]}}.

\bibitem{Sun:2021thf}
Z.~Sun, ``{A note on the representations of $\text{SO}(1,d+1)$},''
  \href{http://arxiv.org/abs/2111.04591}{{\ttfamily arXiv:2111.04591
  [hep-th]}}.

\bibitem{Sengor:2022lyv}
G.~Seng\"or, ``{The de Sitter group and its presence at the late-time
  boundary},'' \href{http://dx.doi.org/10.22323/1.406.0356}{{\em PoS}
  {\bfseries CORFU2021} (2022) 356},
  \href{http://arxiv.org/abs/2206.04719}{{\ttfamily arXiv:2206.04719
  [hep-th]}}.

\bibitem{Sengor:2022kji}
G.~\c{S}eng\"or, ``{Particles of a de Sitter Universe},''
  \href{http://dx.doi.org/10.3390/universe9020059}{{\em Universe} {\bfseries 9}
  no.~2, (2023) 59}, \href{http://arxiv.org/abs/2212.10626}{{\ttfamily
  arXiv:2212.10626 [hep-th]}}.

\bibitem{Enayati:2022hed}
M.~Enayati, J.-P. Gazeau, H.~Pejhan, and A.~Wang,
  \href{http://dx.doi.org/10.1007/978-3-031-16045-5}{{\em {The de Sitter (dS)
  Group and its Representations. An Introduction to Elementary Systems and
  Modeling the Dark Energy Universe}}}.
\newblock Synthesis Lectures on Mathematics \& Statistics. Springer, 2023.
\newblock \href{http://arxiv.org/abs/2201.11457}{{\ttfamily arXiv:2201.11457
  [math-ph]}}.

\bibitem{RiosFukelman:2023mgq}
A.~Rios~Fukelman, M.~Semp\'e, and G.~A. Silva, ``{Notes on gauge fields and
  discrete series representations in de Sitter spacetimes},''
  \href{http://dx.doi.org/10.1007/JHEP01(2024)011}{{\em JHEP} {\bfseries 01}
  (2024) 011}, \href{http://arxiv.org/abs/2310.14955}{{\ttfamily
  arXiv:2310.14955 [hep-th]}}.

\bibitem{Schaub:2024rnl}
V.~Schaub, ``{A Walk Through $Spin(1,d+1)$},''
  \href{http://arxiv.org/abs/2405.01659}{{\ttfamily arXiv:2405.01659
  [hep-th]}}.

\bibitem{Deser:1983mm}
S.~Deser and R.~I. Nepomechie, ``{Gauge Invariance Versus Masslessness in De
  Sitter Space},'' \href{http://dx.doi.org/10.1016/0003-4916(84)90156-8}{{\em
  Annals Phys.} {\bfseries 154} (1984) 396}.

\bibitem{Deser:2001us}
S.~Deser and A.~Waldron, ``{Partial masslessness of higher spins in (A)dS},''
  \href{http://dx.doi.org/10.1016/S0550-3213(01)00212-7}{{\em Nucl. Phys. B}
  {\bfseries 607} (2001) 577--604},
  \href{http://arxiv.org/abs/hep-th/0103198}{{\ttfamily arXiv:hep-th/0103198}}.

\bibitem{Deser:2001pe}
S.~Deser and A.~Waldron, ``{Gauge invariances and phases of massive higher
  spins in (A)dS},''
  \href{http://dx.doi.org/10.1103/PhysRevLett.87.031601}{{\em Phys. Rev. Lett.}
  {\bfseries 87} (2001) 031601},
  \href{http://arxiv.org/abs/hep-th/0102166}{{\ttfamily arXiv:hep-th/0102166}}.

\bibitem{Deser:2003gw}
S.~Deser and A.~Waldron, ``{Arbitrary spin representations in de Sitter from dS
  / CFT with applications to dS supergravity},''
  \href{http://dx.doi.org/10.1016/S0550-3213(03)00348-1}{{\em Nucl. Phys. B}
  {\bfseries 662} (2003) 379--392},
  \href{http://arxiv.org/abs/hep-th/0301068}{{\ttfamily arXiv:hep-th/0301068}}.

\bibitem{DeRham:2018axr}
C.~De~Rham, K.~Hinterbichler, and L.~A. Johnson, ``{On the (A)dS Decoupling
  Limits of Massive Gravity},''
  \href{http://dx.doi.org/10.1007/JHEP09(2018)154}{{\em JHEP} {\bfseries 09}
  (2018) 154}, \href{http://arxiv.org/abs/1807.08754}{{\ttfamily
  arXiv:1807.08754 [hep-th]}}.

\bibitem{Higuchi:1985ad}
A.~Higuchi, ``{Symmetric tensor fields in de sitter space-time},''.

\bibitem{Higuchi:1986py}
A.~Higuchi, ``{Forbidden Mass Range for Spin-2 Field Theory in De Sitter
  Space-time},'' \href{http://dx.doi.org/10.1016/0550-3213(87)90691-2}{{\em
  Nucl. Phys. B} {\bfseries 282} (1987) 397--436}.

\bibitem{Allen:1985ux}
B.~Allen, ``{Vacuum States in de Sitter Space},''
  \href{http://dx.doi.org/10.1103/PhysRevD.32.3136}{{\em Phys. Rev. D}
  {\bfseries 32} (1985) 3136}.

\bibitem{Allen:1987tz}
B.~Allen and A.~Folacci, ``{The Massless Minimally Coupled Scalar Field in De
  Sitter Space},'' \href{http://dx.doi.org/10.1103/PhysRevD.35.3771}{{\em Phys.
  Rev. D} {\bfseries 35} (1987) 3771}.

\bibitem{Vilenkin:1982wt}
A.~Vilenkin and L.~H. Ford, ``{Gravitational Effects upon Cosmological Phase
  Transitions},'' \href{http://dx.doi.org/10.1103/PhysRevD.26.1231}{{\em Phys.
  Rev. D} {\bfseries 26} (1982) 1231}.

\bibitem{Starobinsky:1982ee}
A.~A. Starobinsky, ``{Dynamics of Phase Transition in the New Inflationary
  Universe Scenario and Generation of Perturbations},''
  \href{http://dx.doi.org/10.1016/0370-2693(82)90541-X}{{\em Phys. Lett. B}
  {\bfseries 117} (1982) 175--178}.

\bibitem{Folacci:1992xc}
A.~Folacci, ``{BRST quantization of the massless minimally coupled scalar field
  in de Sitter space: Zero modes, euclideanization and quantization},''
  \href{http://dx.doi.org/10.1103/PhysRevD.46.2553}{{\em Phys. Rev. D}
  {\bfseries 46} (1992) 2553--2559},
  \href{http://arxiv.org/abs/0911.2064}{{\ttfamily arXiv:0911.2064 [gr-qc]}}.

\bibitem{Joung:2007je}
E.~Joung, J.~Mourad, and R.~Parentani, ``{Group theoretical approach to quantum
  fields in de Sitter space. II. The complementary and discrete series},''
  \href{http://dx.doi.org/10.1088/1126-6708/2007/09/030}{{\em JHEP} {\bfseries
  09} (2007) 030}, \href{http://arxiv.org/abs/0707.2907}{{\ttfamily
  arXiv:0707.2907 [hep-th]}}.

\bibitem{Marolf:2008it}
D.~Marolf and I.~Morrison, ``{Group Averaging of massless scalar fields in 1+1
  de Sitter},'' \href{http://dx.doi.org/10.1088/0264-9381/26/3/035001}{{\em
  Class. Quant. Grav.} {\bfseries 26} (2009) 035001},
  \href{http://arxiv.org/abs/0808.2174}{{\ttfamily arXiv:0808.2174 [gr-qc]}}.

\bibitem{Bros:2010wa}
J.~Bros, H.~Epstein, and U.~Moschella, ``{Scalar tachyons in the de Sitter
  universe},'' \href{http://dx.doi.org/10.1007/s11005-010-0406-4}{{\em Lett.
  Math. Phys.} {\bfseries 93} (2010) 203--211},
  \href{http://arxiv.org/abs/1003.1396}{{\ttfamily arXiv:1003.1396 [hep-th]}}.

\bibitem{Epstein:2014jaa}
H.~Epstein and U.~Moschella, ``{de Sitter tachyons and related topics},''
  \href{http://dx.doi.org/10.1007/s00220-015-2308-x}{{\em Commun. Math. Phys.}
  {\bfseries 336} no.~1, (2015) 381--430},
  \href{http://arxiv.org/abs/1403.3319}{{\ttfamily arXiv:1403.3319 [hep-th]}}.

\bibitem{Sengor:2023buj}
G.~\c{S}eng\"or, ``{Searching for discrete series representations at the
  late-time boundary of de Sitter},'' in {\em {15th International Workshop on
  Lie Theory and Its Applications in Physics}}.
\newblock 12, 2023.
\newblock \href{http://arxiv.org/abs/2312.00363}{{\ttfamily arXiv:2312.00363
  [hep-th]}}.

\bibitem{Antoniadis:1991fa}
I.~Antoniadis and E.~Mottola, ``{4-D quantum gravity in the conformal
  sector},'' \href{http://dx.doi.org/10.1103/PhysRevD.45.2013}{{\em Phys. Rev.
  D} {\bfseries 45} (1992) 2013--2025}.

\bibitem{Folacci:1996dv}
A.~Folacci, ``{Toy model for the zero mode problem in the conformal sector of
  de Sitter quantum gravity},''
  \href{http://dx.doi.org/10.1103/PhysRevD.53.3108}{{\em Phys. Rev. D}
  {\bfseries 53} (1996) 3108--3117}.

\bibitem{Faizal:2011iv}
M.~Faizal and A.~Higuchi, ``{Physical equivalence between the covariant and
  physical graviton two-point functions in de Sitter spacetime},''
  \href{http://dx.doi.org/10.1103/PhysRevD.85.124021}{{\em Phys. Rev. D}
  {\bfseries 85} (2012) 124021},
  \href{http://arxiv.org/abs/1107.0395}{{\ttfamily arXiv:1107.0395 [gr-qc]}}.

\bibitem{Mora:2012zi}
P.~J. Mora, N.~C. Tsamis, and R.~P. Woodard, ``{Graviton Propagator in a
  General Invariant Gauge on de Sitter},''
  \href{http://dx.doi.org/10.1063/1.4764882}{{\em J. Math. Phys.} {\bfseries
  53} (2012) 122502}, \href{http://arxiv.org/abs/1205.4468}{{\ttfamily
  arXiv:1205.4468 [gr-qc]}}.

\bibitem{Morrison:2013rqa}
I.~A. Morrison, ``{On cosmic hair and ''de Sitter breaking'' in linearized
  quantum gravity},'' \href{http://arxiv.org/abs/1302.1860}{{\ttfamily
  arXiv:1302.1860 [gr-qc]}}.

\bibitem{Miao:2013isa}
S.~P. Miao, P.~J. Mora, N.~C. Tsamis, and R.~P. Woodard, ``{Perils of analytic
  continuation},'' \href{http://dx.doi.org/10.1103/PhysRevD.89.104004}{{\em
  Phys. Rev. D} {\bfseries 89} no.~10, (2014) 104004},
  \href{http://arxiv.org/abs/1306.5410}{{\ttfamily arXiv:1306.5410 [gr-qc]}}.

\bibitem{Anninos:2021ene}
D.~Anninos, T.~Bautista, and B.~M\"uhlmann, ``{The two-sphere partition
  function in two-dimensional quantum gravity},''
  \href{http://dx.doi.org/10.1007/JHEP09(2021)116}{{\em JHEP} {\bfseries 09}
  (2021) 116}, \href{http://arxiv.org/abs/2106.01665}{{\ttfamily
  arXiv:2106.01665 [hep-th]}}.

\bibitem{Anninos:2023lin}
D.~Anninos, T.~Anous, B.~Pethybridge, and G.~\c{S}eng\"or, ``{The discreet
  charm of the discrete series in dS$_{2}$},''
  \href{http://dx.doi.org/10.1088/1751-8121/ad14ad}{{\em J. Phys. A} {\bfseries
  57} no.~2, (2024) 025401}, \href{http://arxiv.org/abs/2307.15832}{{\ttfamily
  arXiv:2307.15832 [hep-th]}}.

\bibitem{Isler:1989hq}
K.~Isler and C.~A. Trugenberger, ``{A Gauge Theory of Two-dimensional Quantum
  Gravity},'' \href{http://dx.doi.org/10.1103/PhysRevLett.63.834}{{\em Phys.
  Rev. Lett.} {\bfseries 63} (1989) 834}.

\bibitem{Chamseddine:1989yz}
A.~H. Chamseddine and D.~Wyler, ``{Gauge Theory of Topological Gravity in
  (1+1)-Dimensions},''
  \href{http://dx.doi.org/10.1016/0370-2693(89)90528-5}{{\em Phys. Lett. B}
  {\bfseries 228} (1989) 75--78}.

\bibitem{deBoer:2015kda}
J.~de~Boer, M.~P. Heller, R.~C. Myers, and Y.~Neiman, ``{Holographic de Sitter
  Geometry from Entanglement in Conformal Field Theory},''
  \href{http://dx.doi.org/10.1103/PhysRevLett.116.061602}{{\em Phys. Rev.
  Lett.} {\bfseries 116} no.~6, (2016) 061602},
  \href{http://arxiv.org/abs/1509.00113}{{\ttfamily arXiv:1509.00113
  [hep-th]}}.

\bibitem{deBoer:2016pqk}
J.~de~Boer, F.~M. Haehl, M.~P. Heller, and R.~C. Myers, ``{Entanglement,
  holography and causal diamonds},''
  \href{http://dx.doi.org/10.1007/JHEP08(2016)162}{{\em JHEP} {\bfseries 08}
  (2016) 162}, \href{http://arxiv.org/abs/1606.03307}{{\ttfamily
  arXiv:1606.03307 [hep-th]}}.

\bibitem{Farnsworth:2024iwc}
K.~Farnsworth, K.~Hinterbichler, and O.~Hulik, ``{Scale and conformal
  invariance on (A)dS spacetimes},''
  \href{http://dx.doi.org/10.1103/PhysRevD.110.045011}{{\em Phys. Rev. D}
  {\bfseries 110} no.~4, (2024) 045011},
  \href{http://arxiv.org/abs/2402.12430}{{\ttfamily arXiv:2402.12430
  [hep-th]}}.

\bibitem{Farnsworth:2021zgj}
K.~Farnsworth, K.~Hinterbichler, and O.~Hulik, ``{Scale versus conformal
  invariance at the IR fixed point of quantum gravity},''
  \href{http://dx.doi.org/10.1103/PhysRevD.105.066026}{{\em Phys. Rev. D}
  {\bfseries 105} no.~6, (2022) 066026},
  \href{http://arxiv.org/abs/2110.10160}{{\ttfamily arXiv:2110.10160
  [hep-th]}}.

\bibitem{El-Showk:2011xbs}
S.~El-Showk, Y.~Nakayama, and S.~Rychkov, ``{What Maxwell Theory in
  D\ensuremath{<}\ensuremath{>}4 teaches us about scale and conformal
  invariance},'' \href{http://dx.doi.org/10.1016/j.nuclphysb.2011.03.008}{{\em
  Nucl. Phys. B} {\bfseries 848} (2011) 578--593},
  \href{http://arxiv.org/abs/1101.5385}{{\ttfamily arXiv:1101.5385 [hep-th]}}.

\bibitem{Hinterbichler:2024vyv}
K.~Hinterbichler, ``{Dualities among massive, partially massless and shift
  symmetric fields on (A)dS},''
  \href{http://dx.doi.org/10.1007/JHEP06(2024)033}{{\em JHEP} {\bfseries 06}
  (2024) 033}, \href{http://arxiv.org/abs/2402.16938}{{\ttfamily
  arXiv:2402.16938 [hep-th]}}.

\bibitem{Kitaev:2017hnr}
A.~Kitaev, ``{Notes on $\widetilde{\mathrm{SL}}(2,\mathbb{R})$
  representations},'' \href{http://arxiv.org/abs/1711.08169}{{\ttfamily
  arXiv:1711.08169 [hep-th]}}.

\bibitem{Chen:2022wpa}
H.~Chen, ``{Supertranslation Goldstone and de Sitter Tachyons},''
  \href{http://arxiv.org/abs/2212.09646}{{\ttfamily arXiv:2212.09646
  [hep-th]}}.

\bibitem{Carroll:2004st}
S.~M. Carroll, \href{http://dx.doi.org/10.1017/9781108770385}{{\em {Spacetime
  and Geometry}: {An Introduction to General Relativity}}}.
\newblock Cambridge University Press, 7, 2019.

\bibitem{Osborn:1993cr}
H.~Osborn and A.~C. Petkou, ``{Implications of conformal invariance in field
  theories for general dimensions},''
  \href{http://dx.doi.org/10.1006/aphy.1994.1045}{{\em Annals Phys.} {\bfseries
  231} (1994) 311--362}, \href{http://arxiv.org/abs/hep-th/9307010}{{\ttfamily
  arXiv:hep-th/9307010}}.

\bibitem{Allen:1985wd}
B.~Allen and T.~Jacobson, ``{Vector Two Point Functions in Maximally Symmetric
  Spaces},'' \href{http://dx.doi.org/10.1007/BF01211169}{{\em Commun. Math.
  Phys.} {\bfseries 103} (1986) 669}.

\bibitem{Huber:2005yg}
T.~Huber and D.~Maitre, ``{HypExp: A Mathematica package for expanding
  hypergeometric functions around integer-valued parameters},''
  \href{http://dx.doi.org/10.1016/j.cpc.2006.01.007}{{\em Comput. Phys.
  Commun.} {\bfseries 175} (2006) 122--144},
  \href{http://arxiv.org/abs/hep-ph/0507094}{{\ttfamily arXiv:hep-ph/0507094}}.

\bibitem{Polchinski:1998rq}
J.~Polchinski, \href{http://dx.doi.org/10.1017/CBO9780511816079}{{\em {String
  theory. Vol. 1: An introduction to the bosonic string}}}.
\newblock Cambridge Monographs on Mathematical Physics. Cambridge University
  Press, 12, 2007.

\bibitem{Loparco:2023rug}
M.~Loparco, J.~Penedones, K.~Salehi~Vaziri, and Z.~Sun, ``{The
  K\"all\'en-Lehmann representation in de Sitter spacetime},''
  \href{http://dx.doi.org/10.1007/JHEP12(2023)159}{{\em JHEP} {\bfseries 12}
  (2023) 159}, \href{http://arxiv.org/abs/2306.00090}{{\ttfamily
  arXiv:2306.00090 [hep-th]}}.

\bibitem{Narain:2014oja}
G.~Narain, ``{Green's function of the Vector fields on DeSitter Background},''
  \href{http://arxiv.org/abs/1408.6193}{{\ttfamily arXiv:1408.6193 [gr-qc]}}.

\bibitem{Belokogne:2016dvd}
A.~Belokogne, A.~Folacci, and J.~Queva, ``{Stueckelberg massive
  electromagnetism in de Sitter and anti\textendash{}de Sitter spacetimes:
  Two-point functions and renormalized stress-energy tensors},''
  \href{http://dx.doi.org/10.1103/PhysRevD.94.105028}{{\em Phys. Rev. D}
  {\bfseries 94} no.~10, (2016) 105028},
  \href{http://arxiv.org/abs/1610.00244}{{\ttfamily arXiv:1610.00244 [gr-qc]}}.

\bibitem{Hogervorst:2021uvp}
M.~Hogervorst, J.~a. Penedones, and K.~S. Vaziri, ``{Towards the
  non-perturbative cosmological bootstrap},''
  \href{http://dx.doi.org/10.1007/JHEP02(2023)162}{{\em JHEP} {\bfseries 02}
  (2023) 162}, \href{http://arxiv.org/abs/2107.13871}{{\ttfamily
  arXiv:2107.13871 [hep-th]}}.

\bibitem{Penedones:2023uqc}
J.~Penedones, K.~Salehi~Vaziri, and Z.~Sun, ``{Hilbert space of Quantum Field
  Theory in de Sitter spacetime},''
  \href{http://arxiv.org/abs/2301.04146}{{\ttfamily arXiv:2301.04146
  [hep-th]}}.

\bibitem{Zwanziger:1970hk}
D.~Zwanziger, ``{Local Lagrangian quantum field theory of electric and magnetic
  charges},'' \href{http://dx.doi.org/10.1103/PhysRevD.3.880}{{\em Phys. Rev.
  D} {\bfseries 3} (1971) 880}.

\bibitem{Siegel:1983es}
W.~Siegel, ``{Manifest Lorentz Invariance Sometimes Requires Nonlinearity},''
  \href{http://dx.doi.org/10.1016/0550-3213(84)90453-X}{{\em Nucl. Phys. B}
  {\bfseries 238} (1984) 307--316}.

\bibitem{Pasti:1995ii}
P.~Pasti, D.~P. Sorokin, and M.~Tonin, ``{Note on manifest Lorentz and general
  coordinate invariance in duality symmetric models},''
  \href{http://dx.doi.org/10.1016/0370-2693(95)00463-U}{{\em Phys. Lett. B}
  {\bfseries 352} (1995) 59--63},
  \href{http://arxiv.org/abs/hep-th/9503182}{{\ttfamily arXiv:hep-th/9503182}}.

\bibitem{Pasti:1995tn}
P.~Pasti, D.~P. Sorokin, and M.~Tonin, ``{Duality symmetric actions with
  manifest space-time symmetries},''
  \href{http://dx.doi.org/10.1103/PhysRevD.52.R4277}{{\em Phys. Rev. D}
  {\bfseries 52} (1995) R4277--R4281},
  \href{http://arxiv.org/abs/hep-th/9506109}{{\ttfamily arXiv:hep-th/9506109}}.

\bibitem{Pasti:1996vs}
P.~Pasti, D.~P. Sorokin, and M.~Tonin, ``{On Lorentz invariant actions for
  chiral p forms},'' \href{http://dx.doi.org/10.1103/PhysRevD.55.6292}{{\em
  Phys. Rev. D} {\bfseries 55} (1997) 6292--6298},
  \href{http://arxiv.org/abs/hep-th/9611100}{{\ttfamily arXiv:hep-th/9611100}}.

\bibitem{Bandos:1997ui}
I.~A. Bandos, K.~Lechner, A.~Nurmagambetov, P.~Pasti, D.~P. Sorokin, and
  M.~Tonin, ``{Covariant action for the superfive-brane of M theory},''
  \href{http://dx.doi.org/10.1103/PhysRevLett.78.4332}{{\em Phys. Rev. Lett.}
  {\bfseries 78} (1997) 4332--4334},
  \href{http://arxiv.org/abs/hep-th/9701149}{{\ttfamily arXiv:hep-th/9701149}}.

\bibitem{Sen:2015nph}
A.~Sen, ``{Covariant Action for Type IIB Supergravity},''
  \href{http://dx.doi.org/10.1007/JHEP07(2016)017}{{\em JHEP} {\bfseries 07}
  (2016) 017}, \href{http://arxiv.org/abs/1511.08220}{{\ttfamily
  arXiv:1511.08220 [hep-th]}}.

\bibitem{Sen:2019qit}
A.~Sen, ``{Self-dual forms: Action, Hamiltonian and Compactification},''
  \href{http://dx.doi.org/10.1088/1751-8121/ab5423}{{\em J. Phys. A} {\bfseries
  53} no.~8, (2020) 084002}, \href{http://arxiv.org/abs/1903.12196}{{\ttfamily
  arXiv:1903.12196 [hep-th]}}.

\bibitem{Mkrtchyan:2019opf}
K.~Mkrtchyan, ``{On Covariant Actions for Chiral $p-$Forms},''
  \href{http://dx.doi.org/10.1007/JHEP12(2019)076}{{\em JHEP} {\bfseries 12}
  (2019) 076}, \href{http://arxiv.org/abs/1908.01789}{{\ttfamily
  arXiv:1908.01789 [hep-th]}}.

\bibitem{Bansal:2021bis}
S.~Bansal, O.~Evnin, and K.~Mkrtchyan, ``{Polynomial Duality-Symmetric
  Lagrangians for Free p-Forms},''
  \href{http://dx.doi.org/10.1140/epjc/s10052-021-09049-0}{{\em Eur. Phys. J.
  C} {\bfseries 81} no.~3, (2021) 257},
  \href{http://arxiv.org/abs/2101.02350}{{\ttfamily arXiv:2101.02350
  [hep-th]}}.

\bibitem{Avetisyan:2022zza}
Z.~Avetisyan, O.~Evnin, and K.~Mkrtchyan, ``{Nonlinear (chiral) p-form
  electrodynamics},'' \href{http://dx.doi.org/10.1007/JHEP08(2022)112}{{\em
  JHEP} {\bfseries 08} (2022) 112},
  \href{http://arxiv.org/abs/2205.02522}{{\ttfamily arXiv:2205.02522
  [hep-th]}}.

\bibitem{Evnin:2022kqn}
O.~Evnin and K.~Mkrtchyan, ``{Three approaches to chiral form interactions},''
  \href{http://dx.doi.org/10.1016/j.difgeo.2023.102016}{{\em Differ. Geom.
  Appl.} {\bfseries 89} (2023) 102016},
  \href{http://arxiv.org/abs/2207.01767}{{\ttfamily arXiv:2207.01767
  [hep-th]}}.

\bibitem{Evnin:2023cdf}
O.~Evnin, E.~Joung, and K.~Mkrtchyan, ``{Democratic Lagrangians from
  topological bulk},''
  \href{http://dx.doi.org/10.1103/PhysRevD.109.066003}{{\em Phys. Rev. D}
  {\bfseries 109} no.~6, (2024) 066003},
  \href{http://arxiv.org/abs/2309.04625}{{\ttfamily arXiv:2309.04625
  [hep-th]}}.

\bibitem{Hull:2023dgp}
C.~M. Hull, ``{Covariant action for self-dual p-form gauge fields in general
  spacetimes},'' \href{http://dx.doi.org/10.1007/JHEP04(2024)011}{{\em JHEP}
  {\bfseries 04} (2024) 011}, \href{http://arxiv.org/abs/2307.04748}{{\ttfamily
  arXiv:2307.04748 [hep-th]}}.

\bibitem{Brust:2016gjy}
C.~Brust and K.~Hinterbichler, ``{Free \ensuremath{\square}$^{k}$ scalar
  conformal field theory},''
  \href{http://dx.doi.org/10.1007/JHEP02(2017)066}{{\em JHEP} {\bfseries 02}
  (2017) 066}, \href{http://arxiv.org/abs/1607.07439}{{\ttfamily
  arXiv:1607.07439 [hep-th]}}.

\bibitem{Baumann:2017jvh}
D.~Baumann, G.~Goon, H.~Lee, and G.~L. Pimentel, ``{Partially Massless Fields
  During Inflation},'' \href{http://dx.doi.org/10.1007/JHEP04(2018)140}{{\em
  JHEP} {\bfseries 04} (2018) 140},
  \href{http://arxiv.org/abs/1712.06624}{{\ttfamily arXiv:1712.06624
  [hep-th]}}.

\bibitem{Fliss:2021ekk}
J.~R. Fliss, ``{Entanglement in the quantum Hall fluid of dipoles},''
  \href{http://dx.doi.org/10.21468/SciPostPhys.11.3.052}{{\em SciPost Phys.}
  {\bfseries 11} no.~3, (2021) 052},
  \href{http://arxiv.org/abs/2105.07448}{{\ttfamily arXiv:2105.07448
  [cond-mat.str-el]}}.

\bibitem{Blauvelt:2022wwa}
E.~Blauvelt, L.~Engelbrecht, and K.~Hinterbichler, ``{Shift Symmetries and
  AdS/CFT},'' \href{http://dx.doi.org/10.1007/JHEP07(2023)103}{{\em JHEP}
  {\bfseries 07} (2023) 103}, \href{http://arxiv.org/abs/2211.02055}{{\ttfamily
  arXiv:2211.02055 [hep-th]}}.

\bibitem{Bonifacio:2021mrf}
J.~Bonifacio, K.~Hinterbichler, A.~Joyce, and D.~Roest, ``{Exceptional scalar
  theories in de Sitter space},''
  \href{http://dx.doi.org/10.1007/JHEP04(2022)128}{{\em JHEP} {\bfseries 04}
  (2022) 128}, \href{http://arxiv.org/abs/2112.12151}{{\ttfamily
  arXiv:2112.12151 [hep-th]}}.

\bibitem{Goon:2011qf}
G.~Goon, K.~Hinterbichler, and M.~Trodden, ``{Symmetries for Galileons and DBI
  scalars on curved space},''
  \href{http://dx.doi.org/10.1088/1475-7516/2011/07/017}{{\em JCAP} {\bfseries
  07} (2011) 017}, \href{http://arxiv.org/abs/1103.5745}{{\ttfamily
  arXiv:1103.5745 [hep-th]}}.

\bibitem{Goon:2011uw}
G.~Goon, K.~Hinterbichler, and M.~Trodden, ``{A New Class of Effective Field
  Theories from Embedded Branes},''
  \href{http://dx.doi.org/10.1103/PhysRevLett.106.231102}{{\em Phys. Rev.
  Lett.} {\bfseries 106} (2011) 231102},
  \href{http://arxiv.org/abs/1103.6029}{{\ttfamily arXiv:1103.6029 [hep-th]}}.

\end{thebibliography}\endgroup

\end{document}